\documentstyle[12pt]{article}
\def\int {\intop \limits}

\def\fnote#1{\footnote}
\begin{document}

\newcommand{\dst}[1]{\displaystyle{#1}}
\newcommand{\barl}{\begin{array}{rl}}
\newcommand{\ball}{\begin{array}{ll}}
\newcommand{\ear}{\end{array}}
\newcommand{\barc}{\begin{array}{c}}
\newcommand{\e}{\mbox{${\bf e}$}}
\newcommand{\J}{\mbox{${\bf J}$}}
\newcommand{\be}{\begin{equation}}
\newcommand{\ee}{\end{equation}}
\newcommand{\aq}[1]{\label{#1}}
\renewcommand \theequation{\thesection.\arabic{equation}}

\title{Influence of a medium on pair photoproduction \\
and bremsstrahlung}
\author{V. N. Baier
and V. M. Katkov\\
Budker Institute of Nuclear Physics\\ 630090 Novosibirsk,
Russia}

\maketitle

\begin{abstract}

The creation of electron-positron pair by a photon and
the bremsstrahlung of an electron in a medium are considered
in high-energy region, where influence of the multiple scattering on
the processes (the Landau-Pomeranchuk-Migdal (LPM) effect)
becomes essential.
The pair photoproduction probability is calculated with an accuracy
up to the "next to leading logarithm".
The integral characteristics: the radiation length and the total
probability of pair photoproduction are analyzed under influence of
the LPM effect, and the asymptotic expansions of these characteristics
are derived.

\end{abstract}

\newpage
\section{Introduction}

When a charged particle is moving in a medium it scatters on atoms.
With probability $\sim \alpha$ this scattering is accompanied by a radiation.
At high energy the radiation process occurs over a rather long distance,
known as the {\em formation length} $l_c$:
\begin{equation}
l_c=\frac{l_0}{1+\gamma^2 \vartheta_c^2},\quad
l_0=\frac{2\varepsilon \varepsilon'}{m^2\omega},
\label{1.1a}
\end{equation}
where $\omega$ is the energy of emitted photon,
$\varepsilon (m)$ is the energy
(the mass) of a particle, $\varepsilon'=\varepsilon-\omega$,
$\vartheta_c$ is the characteristic angle of photon emission,
the system $\hbar=c=1$ is used.
The spectral distribution of the radiation probability per unit time
inside the thick target (the boundary effects are neglected) can be obtained
from the general formula for the spectral probability derived in the
framework of the operator quasiclassical method (see Eqs.(4.2)-(4.8)
in \cite{BKS}). It can be estimated as
\begin{equation}
dW \sim \frac{\alpha}{\pi l_c}
\frac{\Delta^2(l_c)}{m^2+\varepsilon^2\vartheta_c^2}\frac{d\omega}{\omega},
\label{1.1b}
\end{equation}
where $\alpha=e^2=1/137$, $\Delta^2(l_c)$ is the mean square of
momentum transfer to a projectile from a medium (or an external field)
on the formation length $l_c$.

If the angle of multiple scattering on
the formation length $\vartheta_s \equiv \sqrt{\dot{\vartheta}_s^2l_c}$
is small comparing with the angle $1/\gamma$~
($\gamma=\varepsilon/m$ is the Lorentz factor),
then one can consider scattering as a perturbation and perform the
decomposition over "the potential" of a medium.
The radiation probability in this case
is the incoherent sum of the radiation probabilities on isolated atoms of
a medium defined by the Bethe-Heitler formula. One get from (\ref{1.1b})
for the spectral probability of radiation
per unit time in the case $\vartheta_s \ll 1/\gamma$
($\vartheta_c=1/\gamma,~\Delta^2=\varepsilon^2\vartheta_s^2 \ll m^2$)
\begin{equation}
dW \sim \frac{\alpha}{2\pi l_c} \vartheta_s^2 \gamma^2\frac{d\omega}{\omega}
=\frac{\alpha}{2\pi} \dot{\vartheta}_s^2 \gamma^2 \frac{d\omega}{\omega}.
\label{1.1}
\end{equation}

At an ultrahigh energy it is possible that $\vartheta_s \gg 1/\gamma$.
In this case the characteristic radiation angle (giving the main contribution
into the spectral probability) is defined by the angle of multiple
scattering $\vartheta_s$. The self-consistency condition is
\begin{equation}
\vartheta_c^2=\vartheta_s^2= \dot{\vartheta}_s^2l_c \gg \frac{1}{\gamma^2}.
\label{1.2}
\end{equation}
From the condition (\ref{1.2}) we find
\begin{equation}
\gamma^2 \vartheta_c^4 \simeq \dot{\vartheta}_s^2l_0,
\quad l_c \simeq \frac{l_0}{\gamma^2 \vartheta_c^2} \simeq \frac{1}{\gamma}
\sqrt{\frac{l_0}{\dot{\vartheta}_s^2}}.
\label{1.3}
\end{equation}
In this case one get from (\ref{1.1b}) for
the estimate of the spectral radiation probability
per unit time
\begin{eqnarray}
&& dW \sim \frac{\alpha}{\pi l_c} \frac{d\omega}{\omega}
=\frac{\alpha}{\pi} \frac{\dot{\vartheta}_s^2 \gamma^2}{\nu_0}
\frac{d\omega}{\omega},
\nonumber \\
&&\nu_0^2=\dot{\vartheta}_s^2 \gamma^2 l_0=\frac{4\pi Z^2 \alpha^2 n_a}{m^2}
L l_0 \gg 1,\quad L=\ln\left[\frac{a_s^2}{\lambda_c^2}\left(1+
\gamma^2 \vartheta_c^2 \right)\right],
\label{1.4}
\end{eqnarray}
where $Z$ is the charge of the nucleus, $n_a$ is the number density
of atoms in the medium, $\lambda_c=1/m=(\hbar/mc)$ is the electron Compton
wavelength, $a_s$ is the screening radius of the atom.

So, the formula (\ref{1.1b}) gives the general
description of the radiation process
in terms of the mean momentum transfer valid both in a medium and in an
external field, while formulas (\ref{1.1}) and
(\ref{1.4}) describe the process probability in the particular regimes
in a medium.

Landau and Pomeranchuk were the first who showed that if the formation
length of bremsstrahlung becomes comparable to the distance over which
the multiple scattering becomes important, the bremsstrahlung will be
suppressed \cite{1}. Migdal \cite{M1} developed the quantitative
theory of this phenomenon.

New activity with the theory of the LPM effect
(see \cite{BD}, \cite{BDM}, \cite{Z})
is connected with a very
successful series of experiments performed
at SLAC recently (see \cite{E1}, \cite{E2}).
In these experiments the cross section
of the bremsstrahlung of soft photons with energy from 200~keV to
500~MeV from electrons with energy 8~GeV and 25~GeV is measured
with an accuracy of the order of a few percent. Both LPM and dielectric
suppression are observed and investigated. These experiments were the
challenge for the theory since in all the mentioned papers calculations
are performed to logarithmic accuracy which is not enough for description
of the new experiment. The contribution of the Coulomb corrections (at least
for heavy elements) is larger then experimental errors and these corrections
should be taken into account.

Authors developed the new approach to the theory of the
Landau-Pomeranchuk-Migdal (LPM) effect \cite{L1}
in which the cross section of the bremsstrahlung process
in the photon energies region where the influence of the LPM is very strong
was calculated with a term $\propto 1/L$ , where $L$
is characteristic logarithm of the problem,
and with the Coulomb corrections
taken into account. In the photon energy region, where the LPM effect
is "turned off", the obtained cross section
gives the exact Bethe-Heitler cross section (within power accuracy) with
the Coulomb corrections. This important feature was absent in
the previous calculations.
Some important features of the LPM effect were considered also in \cite{L2},
\cite{L3}, \cite{L4}.

The crossing process for the bremsstrahlung is the pair creation by a
photon. The created particles undergo here the multiple scattering.
It should be emphasize that for the bremsstrahlung the formation length
(\ref{1.2}) increases strongly if $\omega \ll \varepsilon$. Just because
of this the LPM effect was investigated at SLAC at a relatively
low energy. For the pair creation
the formation length $\displaystyle{l_{p}
=\frac{2\varepsilon(\omega-\varepsilon)}{m^2\omega}}$ attains
maximum at $\varepsilon=\omega/2$ and this maximum is
$l_{p,max}=(\omega/2m)\lambda_c$. Because of this even for heavy
elements the effect of
multiple scattering becomes noticeable starting from $\omega \sim 10$~TeV.
Nevertheless it is evident that one have to take into account
the influence of a medium on the pair creation
and on the bremsstrahlung hard part of the spectrum in electromagnetic
showers being created by the cosmic ray particles of the ultrahigh energies.
These effects can be quite significant in the electromagnetic calorimeters
operating in the detectors on the colliders in TeV range.

In the present paper both the spectral probability and
the integral probability of the pair creation are
calculated within an accuracy up to "the next to logarithm"
and with the Coulomb correction taken into account (Sec.2). In Sec.3
the radiation length is calculated under influence of the LPM effect.
The total probability of photon radiation is considered also.
In the Appendixes the technical details of calculation are given.

\section{Influence of multiple scattering on pair creation process}
\setcounter{equation}{0}

The probability of the pair creation by a photon can be obtained
from the probability of the bremsstrahlung with help of the substitution law:
\begin{equation}
\omega^2d\omega \rightarrow \varepsilon^2 d\varepsilon,\quad
\omega \rightarrow -\omega,\quad \varepsilon \rightarrow -\varepsilon,
\label{2.1}
\end{equation}
where $\omega$ is the photon energy, $\varepsilon$ is the energy of
the particle. Making this substitution in Eq.(2.12) of \cite{L1}
we obtain the spectral distribution of the pair creation probability
(over the energy of the created electron)
\begin{equation}
\frac{dW_p}{d\varepsilon}=\frac{2\alpha m^2}{\varepsilon \varepsilon'}
{\rm Im}\left<0|s_1\left(G^{-1}-G_0^{-1}\right)+
s_2 {\bf p}\left(G^{-1}-G_0^{-1}\right){\bf p}|0\right>,
\label{2.2}
\end{equation}
where
\begin{eqnarray}
&& \hspace{-4mm} s_1=1,\quad s_2=\frac{\varepsilon^2+\varepsilon'^2}
{\omega^2},\quad\varepsilon'=\omega-\varepsilon;
\nonumber \\
&& \hspace{-4mm}G_0={\cal H}_0+1,\quad {\cal H}_0={\bf p}^2,\quad
{\bf p}=-i\mbox{\boldmath$\nabla$}_{\mbox{\boldmath$\varrho$}},\quad
G={\cal H}+1,\quad {\cal H}={\bf p}^2-iV(\mbox{\boldmath$\varrho$}),
\quad
\nonumber \\
&& \hspace{-4mm}V(\mbox{\boldmath$\varrho$})=Q\mbox{\boldmath$\varrho$}^2
\left(L_1+\ln \frac{4}{\mbox{\boldmath$\varrho$}^2}-2C \right), \quad
Q=\frac{2\pi Z^2\alpha^2\varepsilon \varepsilon' n_a}{m^4\omega},\quad
L_1=\ln \frac{a_{s2}^2}{\lambda_c^2},
\nonumber \\
&& \hspace{-4mm}\frac{a_{s2}}{\lambda_c}=183Z^{-1/3}{\rm e}^{-f},\quad
f=f(Z\alpha)=(Z\alpha)^2\sum_{k=1}^{\infty}\frac{1}{k(k^2+(Z\alpha)^2)},
\label{2.3}
\end{eqnarray}
where $C=0.577216 \ldots$ is Euler's constant,
$n_a$ is the number density of atoms in the medium,
$\mbox{\boldmath$\varrho$}$ is the coordinate in the two-dimensional space
measured in the Compton wavelength $\lambda_c$, which is conjugate
to the space of the transverse momentum transfers measured in
the electron mass $m$. The mean value in (\ref{2.2}) is taken
over the states with the definite value of the operator
$\mbox{\boldmath$\varrho$}$ (see \cite{L1}, Sec.2).
The contribution of scattering of the created electron and positron on
the atomic electrons can be incorporated into the effective potential
$V(\mbox{\boldmath$\varrho$})$ by substitution
\begin{equation}
\displaystyle{Q \rightarrow Q_{ef}=Q(1+\frac{1}{Z}),\quad
a_{s2} \rightarrow a_{ef}
= a_{s2} \exp \left[\frac{1.88+f(Z \alpha)}{1+Z}\right]}.
\label{2.4}
\end{equation}

The potential $V(\mbox{\boldmath$\varrho$})$ in Eq.(\ref{2.3})
we write in the form
\begin{eqnarray}
&& V(\mbox{\boldmath$\varrho$})=V_c(\mbox{\boldmath$\varrho$})
+v(\mbox{\boldmath$\varrho$}),\quad V_c(\mbox{\boldmath$\varrho$})=
q\mbox{\boldmath$\varrho$}^2, \quad q=QL_c,
\nonumber \\
&&L_c \equiv L(\varrho_c)
=\ln \frac{a_{s2}^2}{\lambda_c^2 \varrho_c^2},\quad
v(\mbox{\boldmath$\varrho$})=-\frac{q\mbox{\boldmath$\varrho$}^2}{L_c}
\left(\ln \frac{\mbox{\boldmath$\varrho$}^2}{4\varrho_c^2}+2C \right),
\label{2.5}
\end{eqnarray}
where the parameter $\varrho_c$ is defined by the set of equations:
\begin{equation}
\varrho_c =1\quad {\rm for}\quad 4QL_1 \leq 1;\quad
4Q\varrho_c^4 \ln \frac{a_{s2}^2}{\lambda_c^2 \varrho_c^2}=1
\quad {\rm for} \quad 4QL_1 \geq 1,
\label{2.6}
\end{equation}
where $L_1$ is defined in Eq.(\ref{2.3}).
The parameter $\varrho_c \simeq 1/p_c$ is determined by the characteristic
angles of created particles with respect to the initial photon momentum
(or the corresponding momentum transfers).
In accordance with such division of the potential we present the propagators
in the expression (\ref{2.2}) as
\begin{equation}
G^{-1}-G_0^{-1}=G^{-1}-G_c^{-1} + G_c^{-1}-G_0^{-1}
\label{2.7}
\end{equation}
where
\[
G_c={\cal H}_c+1,\quad G={\cal H}_c+1-iv,\quad {\cal H}_c={\bf p}^2
-iq \mbox{\boldmath$\varrho$}^2
\]
This representation of the propagator $G^{-1}$ permits one to expand it
over the "perturbation" $v$. Indeed, with an increase of $q$ the relative
value of the perturbation diminishes
$\displaystyle{(\frac{v}{V_c} \sim \frac{1}{L_c})}$ since the
effective impact parameter diminishes and, correspondingly, the value of
logarithm $L_c$ in (\ref{2.5}) increases.
The maximal value of $L_c$ is determined by a size of a nucleus $R_n$
\begin{equation}
L_{max}=\ln \frac{a_{s}^2}{R_n^2} \simeq 2 L_1,
\label{2.8}
\end{equation}
where $a_{s}=a_{s2} \exp (f-1/2) =111Z^{-1/3}\lambda_c$. When
$\varrho_c \ll R_n$ one cannot consider the potential of a nucleus
as the potential of a point charge. In this case the expression
for the potential $V(\mbox{\boldmath$\varrho$})$ has been obtained
in \cite{L1}, Appendix B
\[
V(\mbox{\boldmath$\varrho$})=q\mbox{\boldmath$\varrho$}^2
(L_{max}-0.0407).
\]

The matrix elements of the operator $G_c^{-1}$ was calculated
explicitly in \cite{L1}:
\begin{eqnarray}
&&<\mbox{\boldmath$\varrho$}_1|G_c^{-1}
|\mbox{\boldmath$\varrho$}_2>=i\int_{0}^{\infty}dt e^{-it}
<\mbox{\boldmath$\varrho$}_1|\exp (-i{\cal H}_ct)
|\mbox{\boldmath$\varrho$}_2>,
\nonumber \\
&& <\mbox{\boldmath$\varrho$}_1|\exp (-i{\cal H}_ct)
|\mbox{\boldmath$\varrho$}_2> \equiv
K_c(\mbox{\boldmath$\varrho$}_1, \mbox{\boldmath$\varrho$}_2, t)
\nonumber \\
&& =\frac{\nu}{4\pi i \sinh \nu t} \exp \left\{ \frac{i\nu}{4}
\left[ (\mbox{\boldmath$\varrho$}_1^2+\mbox{\boldmath$\varrho$}_2^2)
\coth \nu t - \frac{2}{\sinh \nu t}
\mbox{\boldmath$\varrho$}_1\mbox{\boldmath$\varrho$}_2\right] \right\},
\label{2.9}
\end{eqnarray}
where $\displaystyle{\nu=2\sqrt{iq}}$. Substituting
this expression in the formula
for the spectral distribution of the pair creation probability (\ref{2.2})
we have
\begin{eqnarray}
&&\displaystyle{\frac{dW_p^c}{d\varepsilon}=
\frac{\alpha m^2}{2\pi \varepsilon \varepsilon'} {\rm Im}~\Phi_p (\nu)},
\nonumber \\
&&\displaystyle{\Phi_p(\nu)=\nu\int_{0}^{\infty} dt e^{-it}\left[s_1
\left(\frac{1}{\sinh z}-\frac{1}{z}\right)-i\nu s_2
\left( \frac{1}{\sinh^2z}- \frac{1}{z^2}\right) \right]}
\nonumber \\
&&=s_1\left(\ln p-\psi\left(p+\frac{1}{2}\right) \right)
+s_2\left(\psi (p) -\ln p+\frac{1}{2p}\right),
\label{2.10}
\end{eqnarray}
where $z=\nu t,~p=i/(2\nu),~\psi(x)$ is the logarithmic derivative of
the gamma function. Some details of the derivation of the last line
can be found in Appendix A (see (\ref{A.1})-(\ref{A.8})).
This formula gives
the spectral distribution of the pair creation probability in the
logarithmic approximation which was used also by
Migdal \cite{M1}. It should be noted that the parameter $\varrho_c$ entering
into the parameter $\nu$ (see Eqs.(\ref{2.3}) and (\ref{2.5}))
is defined up to the factor $\sim 1$, what is inherent in the logarithmic
approximation. However, below we will calculate the next term of
the decomposition over $v(\mbox{\boldmath$\varrho$})$ (an accuracy
up to the "next to leading logarithm") and this permits to obtain
the result which is independent of the parameter $\varrho_c$. It will
be shown that the definition of the parameter $\varrho_c$ in Eq.(\ref{2.6})
minimizes corrections to (\ref{2.10}) practically for all values of
the parameter $\varrho_c$.
It should be emphasized also that here the Coulomb corrections are
included into the parameter $\nu$ in contrast to \cite{M1}.

Let us expand the expression $G^{-1}-G_c^{-1}$ over powers of $v$
\begin{equation}
G^{-1}-G_c^{-1}=G_c^{-1}(iv)G_c^{-1}+G_c^{-1}(iv)G_c^{-1}(iv)G_c^{-1}
+...
\label{2.11}
\end{equation}
Substituting this expansion in (\ref{2.6}) and then in (\ref{2.2})
we obtain the decomposition of the probability of the pair creation.
Let us note that
for $q \ll 1$ the sum of the probability of the pair creation
$\displaystyle{\frac{dW_p^c}{d\varepsilon}}$ (\ref{2.10}) and
the first term of the expansion (\ref{2.11}) gives the Bethe-Heitler spectrum
of electron of created pair, see below (\ref{2.22}).
At $q \geq 1$ the expansion
(\ref{2.11}) is the series over powers of $\displaystyle{\frac{1}{L_c}}$.
It is important that the variation of the parameter $\varrho_c$ by a factor
order of 1 has an influence on the dropped terms
in (\ref{2.11}) only.

In accordance with (\ref{2.7}) and (\ref{2.11}) we present the
probability of radiation in the form
\begin{equation}
\frac{dW_p}{d\varepsilon}=\frac{dW_p^c}{d\varepsilon}+
\frac{dW_p^1}{d\varepsilon}+\frac{dW_p^2}{d\varepsilon}+...
\label{2.12}
\end{equation}
The probability of pair creation $\displaystyle{\frac{dW_p^c}{d\varepsilon}}$
is defined by
Eq.(\ref{2.10}). In formula (\ref{2.2}) with allowance for (\ref{2.7})
there is the expression
\begin{eqnarray}
&& \hspace{-4mm}\displaystyle{<0|G^{-1}-G_c^{-1}|0>=-i\int_{0}^{\infty}dt_1
\int_{0}^{\infty}dt_2 e^{-i(t_1+t_2)}\int_{}^{}d^2\varrho
K_c(0, \mbox{\boldmath$\varrho$}, t_1)
v(\mbox{\boldmath$\varrho$})K_c(\mbox{\boldmath$\varrho$}, 0, t_2)}
\nonumber \\
&& \hspace{-4mm}\displaystyle{+i\int_{0}^{\infty}dt_1
\int_{0}^{\infty}dt_2\int_{0}^{\infty}dt_3
e^{-i(t_1+t_2+t_3)}\int_{}^{}d^2\varrho_1\int_{}^{}d^2\varrho_2
K_c(0, \mbox{\boldmath$\varrho$}_1, t_1)v(\mbox{\boldmath$\varrho$}_1)
K_c(\mbox{\boldmath$\varrho$}_1, \mbox{\boldmath$\varrho$}_2, t_2)}
\nonumber \\
&& \hspace{-4mm}\displaystyle{\times v(\mbox{\boldmath$\varrho$}_2)
K_c(\mbox{\boldmath$\varrho$}_2, 0, t_3) + ...},
\label{2.13}
\end{eqnarray}
where the matrix element $K_c$ is defined by (\ref{2.9}).
The term $\displaystyle{\frac{dW_p^1}{d\varepsilon}}$ in (\ref{2.12})
corresponds to the first term
(linear in $v$) in (\ref{2.13}). Substituting (\ref{2.9}) we have
\begin{eqnarray}
&& \displaystyle{\frac{dW_p^1}{d\varepsilon}=
-\frac{2\alpha m^2}{\varepsilon \varepsilon'}
{\rm Re} \int_{0}^{\infty}dt_1\int_{0}^{\infty}dt_2
e^{-i(t_1+t_2)}\int_{}^{}d^2\varrho
v(\mbox{\boldmath$\varrho$})
\frac{q^2}{\pi^2\nu^2}\frac{1}{\sinh \nu t_1}\frac{1}{\sinh \nu t_2}}
\nonumber \\
&& \displaystyle{\times\exp\left[-\frac{q\varrho^2}{\nu}\left(\coth \nu t_1+
\coth \nu t_2 \right) \right]
\left[s_1+\frac{4q^2\varrho^2}{\nu^2\sinh \nu t_1\sinh \nu t_2}
s_2 \right]}.
\label{2.14}
\end{eqnarray}
Substituting in (\ref{2.14})
the explicit expression for $v(\mbox{\boldmath$\varrho$})$ and
integrating over $d^2\varrho$ and $d(t_1-t_2)$ we obtain
the following formula for the first correction
to the pair creation probability
\begin{eqnarray}
&& \frac{dW_p^1}{d\varepsilon}=
-\frac{\alpha m^2}{4\pi \varepsilon \varepsilon' L}
{\rm Im}~F_p(\nu);\quad F_p(\nu)= \int_{0}^{\infty}\frac{dz e^{-it}}{\sinh^2z}
\left[s_1f_1(z)-2is_2f_2(z) \right],
\nonumber \\
&& f_1(z)=\left(\ln \varrho_c^2+\ln \frac{\nu}{i}
-\ln \sinh z-C\right)g(z) - 2\cosh z G(z),
\nonumber \\
&& f_2(z) = \frac{\nu}{\sinh z}
\left(f_1(z)-\frac{g(z)}{2} \right),
\nonumber \\
&& g(z)=z\cosh z - \sinh z, \quad t=t_1+t_2,~z=\nu t
\nonumber \\
&& G(z)=\int_{0}^{z}(1-y\coth y)dy
\nonumber \\
&&\displaystyle{=z-\frac{z^2}{2}-\frac{\pi^2}{12}-
z\ln \left(1-e^{-2z} \right)
+\frac{1}{2}{\rm Li}_2 \left(e^{-2z} \right)},
\label{2.15}
\end{eqnarray}
here ${\rm Li}_2 \left(x \right)$ is the Euler dilogarithm.
Use of the last representation of function $G(z)$ simplifies
the numerical calculation.

As it was said above (see (\ref{2.6}), (\ref{2.8})), $\varrho_c=1$ at
\begin{equation}
|\nu^2|=\nu_1^2=4QL_1 \leq 1 \quad (q=QL_1).
\label{2.16}
\end{equation}
If the parameter $|\nu| > 1$,
the value of $\varrho_c$ is defined from the equation (\ref{2.6}).
Then one has
\begin{equation}
\displaystyle{\ln \varrho_c^2+\ln\frac{\nu}{i}=\frac{1}{2}
\ln (\varrho_c^4 4QL_c)-i\frac{\pi}{4}=-i\frac{\pi}{4}},\quad
\varrho_c^4 4QL_c=1.
\label{2.17}
\end{equation}
So, we have that the factor at $g(z)$ in the expression
for $f_1(z)$ in (\ref{2.15}) can be written in the form
\begin{eqnarray}
&&\displaystyle{(\ln \varrho_c^2+\ln\frac{\nu}{i}-\ln \sinh z-C) \rightarrow
(\ln \nu_1 \vartheta(1-\nu_1)+\ln(\nu_0\varrho_c^2)\vartheta(\nu_1-1)}
\nonumber \\
&&-i\frac{\pi}{4}-\ln \sinh z -C)=\ln \nu_1 \vartheta(1-\nu_1)
-i\frac{\pi}{4}-\ln \sinh z-C
\label{2.18}
\end{eqnarray}
where
\begin{equation}
\nu_0^2 \equiv |\nu|^2=4q= 4 QL(\varrho_c)=
\frac{8\pi n_aZ^2\alpha^2 \varepsilon \varepsilon'}{m^4 \omega}L(\varrho_c),
\label{2.19}
\end{equation}
$\vartheta(x)$ is the Heaviside step function. So, we have two representation
of $|\nu|$ depending on $\varrho_c$: at $\varrho_c = 1$ it is
$|\nu|=\nu_1$ and at $\varrho_c \leq 1$ it is $|\nu|=\nu_0$.

When the scattering of created particles
is weak ($\nu_1 \ll 1$), the main contribution
in (\ref{2.15}) gives the region where $z \ll 1$. Then
\begin{eqnarray}
&&\displaystyle{f_1(z) \simeq -(C + \ln (it))\frac{z^3}{3}+\frac{2}{9}z^3
=\frac{z^3}{3}(\frac{2}{3}-C-\ln (it))},
\nonumber \\
&&\displaystyle{F_p(\nu)=-\frac{1}{9}\nu^2\left(s_2-
s_1 \right),\quad L \rightarrow L_1}.
\label{2.20}
\end{eqnarray}
Substituting the expansion (\ref{C.1}) into Eq.(\ref{2.10})
we find the corresponding asymptotic decomposition
of the function $\Phi_p(\nu)$
\begin{equation}
\displaystyle{\Phi_p(\nu) \simeq s_1\left(\frac{\nu^2}{6}
+\frac{7\nu^4}{60}+\frac{31\nu^6}{126} \right)+
s_2\left(\frac{\nu^2}{3}
+\frac{2\nu^4}{15}+\frac{16\nu^6}{63} \right),
\quad (|\nu| \ll 1)}
\label{2.21}
\end{equation}
Combining the results obtained (\ref{2.20}) and (\ref{2.21}) we obtain
the spectral distribution of the pair creation probability in the case
when the scattering is weak $(|\nu| \ll 1)$
\begin{eqnarray}
&&\displaystyle{\frac{dW_p}{d\varepsilon}=
\frac{dW_p^c}{d\varepsilon}+\frac{dW_p^1}{d\varepsilon}=
\frac{\alpha m^2}{2\pi \varepsilon \varepsilon'}
{\rm Im}~\left[\Phi_p(\nu)-\frac{1}{2L}F_p(\nu)
\right]}
\nonumber \\
&& \displaystyle{=\frac{\alpha m^2}{2\pi \varepsilon \varepsilon'}
\frac{2Q}{3}\left[s_1\left(
L_1\left(1-\frac{31\nu_1^4}{21} \right)-\frac{1}{3} \right)+2s_2
\left(L_1\left(1-\frac{16\nu_1^4}{21} \right)+\frac{1}{6} \right) \right]}
\nonumber \\
&& \displaystyle{=\frac{4Z^2\alpha^3n_a}{3m^2\omega}\Bigg\{
\left(\ln \left(183Z^{-1/3} \right) -f(Z\alpha)\right)
\left(1-\frac{31\nu_1^4}{21} \right)-\frac{1}{6}}
\nonumber \\
&& \displaystyle{+2\frac{\varepsilon^2+\varepsilon'^2}{\omega^2}
\left[\left(\ln \left(183Z^{-1/3} \right)-
f(Z\alpha)\right)\left(1-\frac{16\nu_1^4}{21} \right)+\frac{1}{12}
\right] \Bigg\}},
\label{2.22}
\end{eqnarray}
where $L_1$ is defined in (\ref{2.3}).
Integrating (\ref{2.22}) over $\varepsilon$ we obtain
\begin{equation}
W_p=\frac{28Z^2\alpha^3n_a}{9m^2}\left[\left(\ln(183 Z^{-1/3})-f(Z\alpha)
\right)\left(1-\frac{3312}{2401} \frac{\omega^2}{\omega_0^2} \right)-
\frac{1}{42} \right],
\label{2.23}
\end{equation}
where
\begin{equation}
\omega_0=m\left(2\pi Z^2 \alpha^2 n_a \lambda_c^3 L_1 \right)^{-1}
\label{2.23a}
\end{equation}
Note that in gold $\omega_0=10.5~$TeV. This is just the value of photon
energy starting with the LPM effect becomes essential for the pair
creation process in heavy elements.
If one omits here the terms $\propto \nu_1^4$
and $\propto (\omega/\omega_0)^2$
these expressions coincide with the known Bethe-Heitler
formula for the probability
of pair creation by a high-energy photon in the case of complete
screening (if one neglects the contribution
of atomic electrons) written down within power accuracy
(omitted terms are of
the order of powers of $\displaystyle{\frac{m}{\omega}}$) with the
Coulomb corrections, see e.g. Eqs.(19.4) and (19.17)
in \cite{BKF}.

The pair creation spectral probability $dW/dx$ vs $x=\varepsilon/\omega$ is
shown in Fig.1 for different energies. It is seen that for $\omega=2.5~$TeV
which below $\omega_0$ the difference with the Bethe-Heitler probability is
rather small. When $\omega > \omega_0$ there is significant difference
with the Bethe-Heitler spectrum increasing with $\omega$ growth.
In Fig.1 are shown the curves (thin lines 2,3,4) obtained
in logarithmic approximation $dW_p^c/d\varepsilon$ (\ref{2.10}),
the first correction to the spectral probability $dW_p^1/d\varepsilon$
(\ref{2.15}), curves $c2, c3, c4$ and the sum of these two contributions:
curves $T1, T2, T3, T4$. It should be noted that for our definition of
the parameter $\varrho_c$ (\ref{2.6}) the corrections are not exceed 6\%
of the main term. The corrections are maximal for $\nu_0 \sim 3$.

The total probability of pair creation in the logarithmic approximation
can be presented as (see (\ref{2.10}))
\begin{eqnarray}
&&\frac{W_p^c}{W_{p0}^{BH}}=\frac{9}{14}\frac{\omega_0}{\omega}
{\rm Im}\int_{0}^{1}\frac{dy}{y(1-y)}\Bigg[
\left(\ln p-\psi\left(p+\frac{1}{2}\right) \right)
\nonumber \\
&&+\left(1-2y+2y^2 \right)\left(\psi\left(p\right)
-\ln p+\frac{1}{2p}\right)\Bigg],
\label{2.23b}
\end{eqnarray}
where
\[
p=\frac{bs}{4},\quad s=\frac{1}{\sqrt{y(1-y)}},\quad
b=\exp \left(i\frac{\pi}{4}\right)
\sqrt{\frac{L_1}{L_c}\frac{\omega_0}{\omega}},
\]
$W_{p0}^{BH}$ is the Bethe-Heitler probability of pair photoproduction
in the logarithmic approximation. The total probability of pair creation
$W_p^c$ in gold is given in Fig.2 (curve 2),it reduced by 10\%
at $\omega \simeq 9$~TeV and it cuts in half at
$\omega \simeq 130$~TeV.

At $\nu_0 \gg 1$ the main term of the function $F_p(\nu)$
(see (\ref{2.15}) and (\ref{2.19}))
can be written in the form
\begin{equation}
\displaystyle{F_p(\nu)= \int_{0}^{\infty}\frac{dz}{\sinh^2z}
\left[s_1f_1(z)-2is_2f_2(z) \right]}.
\label{2.24}
\end{equation}
Integrating over $z$ we obtain
\begin{equation}
\displaystyle{-{\rm Im}~F_p(\nu)=\frac{\pi}{4}(s_1-s_2)+\frac{\nu_0}{\sqrt{2}}
\left(\ln 2-C+\frac{\pi}{4} \right)s_2},
\label{2.25}
\end{equation}
where we take into account the next terms of the decomposition in the
term $\propto s_2$.
Under the same conditions ($\nu_0 \gg 1$) the function ${\rm Im}~\Phi_p(\nu)$
(\ref{2.10}) is
\begin{equation}
\displaystyle{{\rm Im}~\Phi_p(\nu)=\frac{\pi}{4}(s_1-s_2)+
\frac{\nu_0}{\sqrt{2}}s_2}.
\label{2.26}
\end{equation}
Thus, at $\nu_0 \gg 1$ the relative contribution of the first correction
$\displaystyle{\frac{dW_p^1}{d\varepsilon}}$ is defined by
\begin{equation}
\displaystyle{r=\frac{dW_p^1}{dW_p^c}=\frac{1}{2L_c}
\left(\ln 2-C+\frac{\pi}{4} \right) \simeq \frac{0.451}{L_c}}.
\label{2.27}
\end{equation}
In this expression the value $r$ with the accuracy up to terms $\sim 1/L_c^2$
doesn't depend on the energy:$L_c \simeq L_1+\ln(\omega/\omega_0)/2$.
Hence we can find the correction to the total probability at
$\omega \gg \omega_0$. The maximal value of the correction is attained at
$\omega \sim 10\omega_0$, it is $\sim 6\%$ for heavy elements.

When the parameter $\nu_0^2$ is not very large ($\nu_0 < 10^3,~
\varrho_c > R_n$, see (\ref{2.8})) one
can solve the equation $\nu_0^2 \varrho_c^4=1$ (\ref{2.17}) using the
method of successive approximations. In the first approximation we have
\begin{eqnarray}
&&\nu_0^2=\nu_1^2 L_c,\quad L_c \simeq L_1 \left(1+
\frac{\ln \nu_1}{L_1}\vartheta(\nu_1-1) \right),
\nonumber \\
&&\nu_0 \simeq \frac{1}{\cosh \xi}\sqrt{\frac{\omega}{\omega_0}}
\left[1+\frac{1}{4L_1}\left(\ln \frac{\omega}{\omega_0}-
2\ln \cosh \xi \right) \right].
\label{2.28}
\end{eqnarray}
It should be noted that the relative error in the expression for $L_c$
at $\varrho_c > R_n$
is less than $\ln 2/(4L_1) \leq 2.5$\%.
Here we introduce variable $\xi$
\begin{equation}
\frac{\varepsilon}{\omega} = \frac{1}{2}\left(1+\tanh \xi \right),\quad
\frac{\varepsilon \varepsilon'}{\omega^2}=\frac{1}{4 \cosh^2 \xi},\quad
-\infty < \xi <\infty.
\label{2.29}
\end{equation}
Substituting the terms $\propto \nu_0$ in the asymptotic
formulas (\ref{2.26}) and (\ref{2.25}) into the
first line of Eq.(\ref{2.22}) we obtain expression which contain
the integral of the type
\begin{equation}
\int_{-\infty}^{\infty} \frac{d\xi}{\cosh \xi}
\left(1-\frac{1}{2 \cosh^2 \xi} \right)
\left[A+B \ln \cosh \xi) \right]
=\frac{3\pi}{4}\left[A+B\left(\ln 2 + \frac{1}{6} \right) \right].
\label{2.30}
\end{equation}
Using this result we obtain the total probability of pair creation
under strong influence of multiple scattering ($\nu_0 \gg 1$, but
not very large)
\begin{eqnarray}
&& W_p \simeq \frac{3\alpha}{4\sqrt{2}} \frac{m^2}{\sqrt{\omega \omega_0}}
\left[1+\frac{1}{4L_1}
\left(\ln \frac{\omega_0}{\omega}+D \right) \right]
\nonumber \\
&& = \frac{3\pi Z^2 \alpha^3 n_a L_1}{2\sqrt{2}m^2}
\sqrt{\frac{\omega_0}{\omega}}\left[1+\frac{1}{4L_1}
\left(\ln \frac{\omega}{\omega_0}+D \right) \right],
\nonumber \\
&& D=\frac{\pi}{2}-2C - \frac{1}{3} \simeq 0.08303 \simeq \frac{1}{12.04}.
\label{2.31}
\end{eqnarray}

It should be noted that only the main term of the decomposition ($\propto
\nu_0$) can be used in Eqs.(\ref{2.25}) and (\ref{2.26}) for the calculation
of the total probability of pair creation. In the interval
$\omega \gg \omega_0$ the contribution into the correction terms gives also
the region where $\cosh^2 \xi \sim \omega/\omega_0$, where the
parameter $\nu_0 \sim 1$ and the expansion used in
Eq.(\ref{2.24}) is ineligible. The next terms (without corrections
$\propto 1/L_1$) are found in Appendix A
(Eq.(\ref{A.12})), so we have
\begin{equation}
 W_p \simeq \frac{3\alpha}{4\sqrt{2}} \frac{m^2}{\sqrt{\omega \omega_0}}
\left[1-\frac{\sqrt{2}}{3}\left(4\ln 2-1 \right)
\sqrt{\frac{\omega_0}{\omega}}-\frac{\pi^2}{18}\frac{\omega_0}{\omega}+
\frac{1}{4L_1}
\left(\ln \frac{\omega}{\omega_0}+D \right) \right].
\label{2.32}
\end{equation}
In terms of the Bethe-Heitler total probability of pair creation this
result is
\begin{equation}
\frac{W_p}{W^{BH}_p}\simeq 2.14 \sqrt{\frac{\omega_0}{\omega}}
\left[1-0.836 \sqrt{\frac{\omega_0}{\omega}}-0.548 \frac{\omega_0}{\omega}
 + \frac{1}{4L_1}
\left(\ln \frac{\omega}{\omega_0}+0.274 \right) \right]
\label{2.33}
\end{equation}

\section{Influence of the multiple scattering
on the bremsstrahlung}
\setcounter{equation}{0}

The spectral radiation intensity obtained in \cite{L1} (see Eq.(2.39))
has the form
\begin{equation}
dI=\omega dW=\frac{\alpha m^2 xdx}{2\pi (1-x)}
{\rm Im}~\left[\Phi(\nu)-\frac{1}{2L_c}F(\nu)
\right],\quad x=\frac{\omega}{\varepsilon},
\label{3.1}
\end{equation}
where
\begin{eqnarray}
&&\displaystyle{\Phi(\nu)=\int_{0}^{\infty} dz e^{-it}\left[r_1
\left(\frac{1}{\sinh z}-\frac{1}{z}\right)-i\nu r_2
\left( \frac{1}{\sinh^2z}- \frac{1}{z^2}\right) \right]}
\nonumber \\
&& = r_1\left(\ln p-\psi\left(p+\frac{1}{2}\right) \right)
+r_2\left(\psi (p) -\ln p+\frac{1}{2p}\right),
\nonumber \\
&&F(\nu)= \int_{0}^{\infty}\frac{dz e^{-it}}{\sinh^2z}
\left[r_1f_1(z)-2ir_2f_2(z) \right],
\nonumber \\
&& t=\frac{z}{\nu},\quad r_1=x^2,\quad r_2=1+(1-x)^2.
\label{3.2}
\end{eqnarray}
where $z=\nu t,~p=i/(2\nu),~\psi(x)$ is the logarithmic derivative of
the gamma function.Some details of the derivation of the second line
can be found in Appendix A (see (\ref{A.1})-(\ref{A.8})).
The functions $f_1(z)$ and $f_2(z)$ are defined by Eq.(\ref{2.15}),
\begin{eqnarray}
&& \nu^2=i\nu_0^2,\quad \nu_0^2=|\nu|^2 \simeq \nu_1^2\left(1+
\frac{\ln \nu_1}{L_1}\vartheta(\nu_1-1) \right),\quad
\nu_1^2=\frac{\varepsilon}{\varepsilon_0}\frac{1-x}{x},
\nonumber \\
&&\varepsilon_0=m\left(8\pi Z^2 \alpha^2 n_a \lambda_c^3 L_1 \right)^{-1},
\quad L_c \simeq L_1 \left(1+
\frac{\ln \nu_1}{L_1}\vartheta(\nu_1-1) \right)
\label{3.3}
\end{eqnarray}
Note, that the parameter $\varepsilon_0$ is four times smaller than
the parameter $\omega_0$ defined in Eq.(\ref{2.23a}).
The LPM effect manifests
itself when
\begin{equation}
\nu_1(x_c)=1,\quad x_c=\frac{\varepsilon}{\varepsilon_0+\varepsilon} .
\label{3.4}
\end{equation}
The formulas derived in \cite{L1} and written down above
are valid for any energy. In Fig.3 the spectral radiation intensity in gold
($\varepsilon_0=2.5$~TeV) is shown for different energies of
the initial electron. In the case when
$\varepsilon \ll \varepsilon_0$ ($\varepsilon=25~$GeV
and $\varepsilon=250~$GeV) the LPM suppression is seen in the soft part
of the spectrum only for $x \leq x_c \simeq \varepsilon/\varepsilon_0 \ll 1$
while in the region
$\varepsilon \geq \varepsilon_0$ ($\varepsilon=2.5~$TeV and
$\varepsilon=25~$TeV)
where $x_c \sim 1$ the LPM effect is significant  for any $x$.
For relatively low energies $\varepsilon=25~$GeV and $\varepsilon=8~$GeV
used in famous SLAC experiment \cite{E1}, \cite{E2}
we have analyzed the
soft part of spectrum, including all the accompanying effects:
the boundary photon emission, the multiphoton radiation and
influence of the polarization of the medium.
The perfect agreement of the theory and data
was achieved in the whole interval of measured photon energies
(200~keV$ \leq \omega \leq$500~MeV),
see the corresponding figures in
\cite{L1},\cite{L2},\cite{L3}. It should be pointed out that
 both the correction term with $F(\nu)$ and the Coulomb corrections
have to be taken into account for this agreement.

In the case $\varepsilon \ll \varepsilon_0$ in the hard part of spectrum
($1 \geq x \gg x_c$) the parameter $\nu_1^2 \simeq x_c/x \ll 1$ and the
contribution into the integral (\ref{3.2}) give the region $z \ll 1$.
Using the decomposition (\ref{C.1}) we find (compare with (\ref{2.21}),
(\ref{2.22}))
\begin{eqnarray}
&&\displaystyle{{\rm Im}~\Phi(\nu) \simeq r_1\frac{\nu_1^2}{6}
\left(1- \frac{31}{21}\nu_1^4\right)+r_2\frac{\nu_1^2}{3}
\left(1- \frac{16}{21}\nu_1^4\right)},
\nonumber \\
&& -{\rm Im}~F(\nu)= -\frac{1}{9}(r_2-r_1)\nu_1^2 (1 + O(\nu_1^4)).
\label{3.5}
\end{eqnarray}
In the last formula, which presents corrections $\sim 1/L_1$ we restricted
ourselves to the main terms of expansion.
Substituting into (\ref{3.1}) we have
\begin{eqnarray}
&& \frac{dI}{dx}=\frac{2Z^2\alpha^3n_a\varepsilon}{3m^2}\Bigg[r_1
\left(L_1\left(1-\frac{31}{21}\frac{x_c^2}{x^2}(1-x)^2 \right)-
\frac{1}{3} \right)
\nonumber \\
&&+2r_2
\left(L_1\left(1-\frac{16}{21}\frac{x_c^2}{x^2}(1-x)^2 \right)+
\frac{1}{6} \right)
\label{3.6}
\end{eqnarray}
Note that if neglect here the terms $\propto x_c^2/x^2$ we obtain
the Bethe-Heitler intensity spectrum with the Coulomb corrections.

In the case $\varepsilon \geq \varepsilon_0$ the intensity spectrum
differs from the Bethe-Heitler one at $x \sim 1$ also. When
$\varepsilon \gg \varepsilon_0$ one can use the asymptotic expansions
(\ref{2.25}) and (\ref{2.26}) in the interval not very close
to the end of the spectrum ($x=1$):
\begin{eqnarray}
&& \frac{dI}{dx}=\frac{\alpha m^2 \nu_0 x}{2 \sqrt{2} \pi (1-x)}\left[
\frac{r_1 \pi}{2 \sqrt{2} \nu_0}+r_2 \left(1-
\frac{\pi}{2 \sqrt{2} \nu_0} \right) +r\right]
\nonumber \\
&& \simeq \frac{2\sqrt{2}Z^2\alpha^3n_a\varepsilon}{m^2}
\sqrt{\frac{\varepsilon_0 x}{\varepsilon (1-x)}}
\left(1+\frac{1}{4L_1}\ln \frac{\varepsilon (1-x)}{\varepsilon_0 x} \right)
\Bigg[x^2\nonumber \\
&& +2(1-x)\left(1-\frac{\pi}{2\sqrt{2}}
\sqrt{\frac{\varepsilon_0 x}{\varepsilon (1-x)}}\right)+r \Bigg],\quad
\varepsilon(1-x) \gg \varepsilon_0x.
\label{3.7}
\end{eqnarray}

Now we turn to the integral characteristics of radiation.
The total intensity of radiation in the logarithmic approximation
can be presented as (see (\ref{3.1}))
\begin{eqnarray}
&&\frac{I}{\varepsilon}L_{rad}^0=2\frac{\varepsilon_0}{\varepsilon}
{\rm Im}\Bigg[\int_{0}^{1}\frac{dx}{g}\sqrt{\frac{x}{1-x}}(2(1-x)+x^2)
\nonumber \\
&&+ \int_{0}^{1}\frac{x^3dx}{1-x}
\left(\psi(p+1)-\psi\left(p+\frac{1}{2}\right) \right)
+2\int_{0}^{1}xdx\left(\psi\left(p+1\right)
-\ln p\right)\Bigg],
\label{3.7a}
\end{eqnarray}
where
\[
p=\frac{g\eta}{2},\quad \eta=\sqrt{\frac{x}{1-x}},\quad
g=\exp \left(i\frac{\pi}{4}\right)
\sqrt{\frac{L_1}{L_c}\frac{\varepsilon_0}{\epsilon}},
\]
$L_{rad}^0$ is the radiation length in the logarithmic approximation.
The relative energy losses of electron per unit time in terms
of the Bethe-Heitler radiation length $L_{rad}^0$:
$\displaystyle{\frac{I}{\varepsilon}L_{rad}^0}$
in gold is given in Fig.2 (curve 1),
it reduces by 10\% (15\% and 25\%)
at $\varepsilon \simeq 700$~GeV ($\varepsilon \simeq 1.4$~TeV and
$\varepsilon \simeq 3.8$~TeV)
respectively, and it cuts in half at $\omega \simeq 26$~TeV.
This increase of effective radiation length can be important
in electromagnetic calorimeters
operating in detectors on colliders in TeV range.
The contribution of the correction terms was discussed after (\ref{2.27}).
It is valid for the radiation process also.

The spectral distribution of bremsstrahlung intensity
and the spectral distribution over energy of created electron (positron)
as well as the reduction of energy loss and the photon conversion
cross section was calculated by Klein \cite{K}, \cite{E2} using the Migdal
\cite{M1} formulas. As was explained above (after Eq.(\ref{2.10}))
we use more accurate procedure of fine tuning and because of this
our calculation in logarithmic approximation differs from Migdal one.
We calculated also the correction term and include the Coulomb
corrections. For this reason the results shown here in Figs.1-3
are more precise than given in \cite{K}, \cite{E2}.

In Eqs.(\ref{3.6}) and (\ref{3.7}) we can use the main terms of
decomposition only. The main term in (\ref{3.6}) gives after the integration
over $x$ the standard expression for the radiation length $L_{rad}$
without influence of multiple scattering. The correction term is calculated
in Appendix C (see (\ref{C.9})) where we need to put
$|\beta|^2=\varepsilon_0/\varepsilon \gg 1$
\begin{eqnarray}
&&\frac{I}{\varepsilon}=\frac{\alpha m^2}{4\pi \varepsilon_0}
\left(1+\frac{1}{9L_1}-\frac{4\pi}{15}\frac{\varepsilon}{\varepsilon_0}
\right) \simeq L_{rad}^{-1}\left(1
-\frac{4\pi}{15}\frac{\varepsilon}{\varepsilon_0}\right),
\nonumber \\
&& \frac{1}{L_{rad}}= \frac{2Z^2\alpha^3n_a L_1}{m^2}
\left(1+\frac{1}{9L_1} \right)
\label{3.8}
\end{eqnarray}

The integration over $x$  of the main term in (\ref{3.7}) gives
(terms $\propto \sqrt{\varepsilon_0/\varepsilon}$ in the square brackets
are neglected)
\begin{eqnarray}
&& I_0 \simeq \frac{9\pi Z^2 \alpha^3 n_a \sqrt{\varepsilon \varepsilon_0}}
{4\sqrt{2}m^2} L_1 \left[1+\frac{1}{4L_1}
\left(\ln \frac{\varepsilon}{\varepsilon_0}-\frac{46}{27} \right)+r_0 \right]
\nonumber \\
&& r_0=\frac{1}{2L_1}\left(\ln 2 -C +\frac{\pi}{4} \right).
\label{3.9}
\end{eqnarray}
The corrections (without terms $\propto 1/L_1$) to (\ref{3.9})
are calculated in Appendix B
(see Eq.(\ref{B.11})). The complete result is
\begin{eqnarray}
&& I = \frac{9\alpha m^2}{32\sqrt{2}}\sqrt{\frac{\varepsilon}{\varepsilon_0}}
\Bigg[1-\frac{4\sqrt{2}}{9}\left(4\ln 2+1 \right)
\sqrt{\frac{\varepsilon_0}{\varepsilon}}
-\frac{25\pi^2}{54}\frac{\varepsilon_0}{\varepsilon}
\nonumber \\
&& +\frac{1}{4L_1}
\left(\ln \frac{\varepsilon}{\varepsilon_0}
+2\ln 2-2C+\frac{\pi}{2}-\frac{46}{27} \right) \Bigg],
\nonumber \\
&& \frac{I}{\varepsilon L_{rad}} \simeq \frac{5}{2}
\sqrt{\frac{\varepsilon_0}{\varepsilon}}
\left[1-2.37\sqrt{\frac{\varepsilon_0}{\varepsilon}}
-4.57 \frac{\varepsilon_0}{\varepsilon}
+\frac{1}{4L_1}
\left(\ln \frac{\varepsilon}{\varepsilon_0}
-0.3455 \right)\right]
\label{3.10}
\end{eqnarray}
Although the coefficients in the last expression are rather large at
two first terms of the decomposition over
$\sqrt{\varepsilon_0/\varepsilon}$ this formula has the accuracy of the order
of 10\% at $\varepsilon \sim 10 \varepsilon_0$.
The integral probability of radiation was calculated in \cite{L3}
\begin{equation}
W=\frac{4}{3L_{rad}}\left(\ln \frac{\varepsilon_0}{\varepsilon}+C_2 \right),
\quad C_2=2C-\frac{5}{8}+\int_{0}^{\infty}\ln z
\left(\frac{1}{z^3}\frac{\cosh z}{\sinh^3 z} \right) \simeq 1.96
\label{3.11}
\end{equation}

In the case $\varepsilon \gg \varepsilon_0$
we can calculate the integral probability of radiation starting with
Eq.(\ref{3.7}). Conserving the main term, dividing it by $x\varepsilon$
and integrating over $x$ we find
\begin{equation}
W_0=\frac{11\pi Z^2 \alpha^3 n_a}{2 \sqrt{2} m^2}
\sqrt{\frac{\varepsilon_0}{\varepsilon}} L_1
\left[1+ \frac{1}{4L_1}
\left(\ln \frac{\varepsilon}{\varepsilon_0}
+\frac{8}{11} \right)+r_0 \right]
\label{3.12}
\end{equation}

The correction terms to Eq.(\ref{3.11}) are calculated in Appendix B
(see Eq.(\ref{B.13})). Substituting we have
\begin{eqnarray}
\hspace{-15mm}&& W = \frac{11\alpha m^2}{16\sqrt{2\varepsilon \varepsilon_0}}
\Bigg[1-\frac{4\sqrt{2}}{11}\left(2\ln 2+1 \right)
\sqrt{\frac{\varepsilon_0}{\varepsilon}}
+\frac{\pi^2}{6}\frac{\varepsilon_0}{\varepsilon}
\nonumber \\
\hspace{-15mm}&& +\frac{1}{4L_1}
\left(\ln \frac{\varepsilon}{\varepsilon_0}
+2\ln 2-2C+\frac{\pi}{2}+\frac{8}{11} \right) \Bigg]
\nonumber \\
\hspace{-15mm}&& =\frac{11\pi Z^2 \alpha^3 n_a}{2\sqrt{2} m^2}
\sqrt{\frac{\varepsilon_0}{\varepsilon}}L_1
\left[1-1.23\sqrt{\frac{\varepsilon_0}{\varepsilon}}
+1.645 \frac{\varepsilon_0}{\varepsilon}
+\frac{1}{4L_1}
\left(\ln \frac{\varepsilon}{\varepsilon_0}
+2.53 \right)\right].
\label{3.13}
\end{eqnarray}

Ratio of the main terms of Eqs.(\ref{3.10}) and (\ref{3.13})
gives the mean energy of radiated photon
\begin{equation}
\bar{\omega}=\frac{9}{22}\varepsilon \simeq 0.409 \varepsilon.
\label{3.14}
\end{equation}

\newpage
\setcounter{equation}{0}
\Alph{equation}
\appendix

\section{Appendix}

We consider the integral which represent the integral probability of pair
photoproduction (see Eqs.(\ref{2.10}) and (\ref{2.29}))
\begin{eqnarray}
&&\Pi(a)=\int_{0}^{\infty}d\xi \int_{0}^{\infty}dz \exp(-az\cosh \xi)
\Bigg[\frac{1}{\sinh z}-\frac{1}{z}
\nonumber \\
&&+\frac{1}{a\cosh \xi}
\left(1-\frac{1}{2\cosh^2 \xi} \right)
\left( \frac{1}{\sinh^2z}- \frac{1}{z^2}\right) \Bigg].
\label{A.1}
\end{eqnarray}
Integrating by parts (over $z$) the second term of the integrand in
(\ref{A.1}) we have
\begin{eqnarray}
&& \Pi(a)=-\frac{1}{a}\int_{0}^{\infty} \frac{d\xi}{\cosh \xi}
\left(1-\frac{1}{2\cosh^2 \xi} \right)+
\int_{0}^{\infty}d\xi \int_{0}^{\infty}dz \exp(-az\cosh \xi)
\Bigg[\frac{1}{\sinh z}
\nonumber \\
&&+1-\coth z- \frac{1}{2\cosh^2 \xi}
\left(1-\coth z+ \frac{1}{z} \right) \Bigg].
\label{A.2}
\end{eqnarray}
The functions entering in (\ref{A.2}) we present as
\begin{equation}
\frac{1}{\sinh z}=2\sum_{k=1}^{\infty} \exp(-(2k-1)z),\quad
\coth z-1=2\sum_{k=1}^{\infty} \exp(-2kz)
\label{A.3}
\end{equation}
Let us consider the integral entering (\ref{A.2})
\begin{eqnarray}
&&\pi_1(a)=\int_{0}^{\infty}\frac{d \xi}{\cosh^2 \xi}F_1(a\cosh \xi),
\nonumber \\
&&F_1(a\cosh \xi)=\int_{0}^{\infty}dz
\exp(-az\cosh \xi)\left(1-\coth z+ \frac{1}{z} \right)
\label{A.4}
\end{eqnarray}
To avoid a divergence of the individual terms in the integral over $z$
we put the lower limit of the integration $\delta \rightarrow 0$. Using
(\ref{A.3}) we obtain
\begin{equation}
F_1(x)= \lim_{\delta \rightarrow 0}
\left[-{\rm Ei}(-\delta x) -\sum_{k=1}^{\infty}
\frac{\exp(-2k\delta)}{k} + \sum_{k=1}^{\infty}
\frac{x}{k(2k+x)}\right]
\label{A.5}
\end{equation}
Using the expansions
\begin{eqnarray}
&&-{\rm Ei}(-\delta x) = -\ln (\delta x)-C,\quad
-\sum_{k=1}^{\infty}\frac{\exp(-2k\delta)}{k} = \ln(1-\exp(-2\delta))
=\ln 2\delta,
\nonumber \\
&& F_1(x)=\psi\left(\frac{x}{2}+1 \right) -\ln \frac{x}{2},
\label{A.6}
\end{eqnarray}
where $\psi(x)$ is the logarithmic derivative of
the gamma function,
and taking integrals over $\xi$ we have
\begin{equation}
\pi_1(a)= \ln \frac{4}{a}-C -1 +\frac{a}{4}\sum_{k=1}^{\infty}\frac{1}{k^2}
-a^2\sum_{k=1}^{\infty}\frac{1}{k^2\sqrt{4k^2-a^2}}
\ln \frac{\sqrt{2k+a}+\sqrt{2k-a}}{\sqrt{2a}}
\label{A.7}
\end{equation}
The formula (\ref{A.2}) contains also the integral
\begin{eqnarray}
\hspace{-15mm}&& \pi_2(a)= \int_{0}^{\infty}d\xi F_2(a\cosh \xi)
\nonumber \\
\hspace{-15mm}&&F_2(x)=
\int_{0}^{\infty}dz \exp(-zx)\left(1-\coth z+
\frac{1}{\sinh z} \right) =\psi\left(\frac{x}{2}+1 \right) -
\psi\left(\frac{x+1}{2} \right)
\label{A.8}
\end{eqnarray}
Transposing the integration order and using (\ref{A.3}) we find
\begin{equation}
\pi_2(a)=2 \sum_{k=1}^{\infty} \int_{0}^{\infty} dz
K_0(az)\left(\exp(-(2k-1)z)-\exp(-2kz) \right),
\label{A.9}
\end{equation}
where $K_0(x)$ is the modified Bessel function. Taking here integrals
we have
\begin{eqnarray}
&&\pi_2(a)= 2\sum_{k=1}^{\infty} \Bigg[
\frac{1}{\sqrt{(2k-1)^2-a^2}}
\ln \frac{2k-1+\sqrt{(2k-1)^2-a^2}}{a}
\nonumber \\
&&-\frac{1}{\sqrt{4k^2-a^2}}
\ln \frac{2k+\sqrt{4k^2-a^2}}{a} \Bigg]
\label{A.10}
\end{eqnarray}
Substituting (\ref{A.7}) and (\ref{A.10}) into Eq.(\ref{A.2}) we obtain
\begin{eqnarray}
&& \Pi(a)= -\frac{3\pi}{8a}+\frac{1}{2}\left(\ln \frac{a}{4}+1+C \right)
-\frac{\pi^3 a}{48}
\nonumber \\
&&+\sum_{k=1}^{\infty}\Bigg[ \frac{a^2}{2k^2\sqrt{4k^2-a^2}}
\ln \frac{\sqrt{2k+a}+\sqrt{2k-a}}{\sqrt{2a}}
\nonumber \\
&&+\frac{2}{\sqrt{(2k-1)^2-a^2}}
\ln \frac{2k-1+\sqrt{(2k-1)^2-a^2}}{a}
\nonumber \\
&&-\frac{2}{\sqrt{4k^2-a^2}}
\ln \frac{2k+\sqrt{4k^2-a^2}}{a} \Bigg]
\label{A.11}
\end{eqnarray}

This expression is particularly convenient at $|a| \leq 1$. In the case
$|a| \ll 1$ the first three terms of the decomposition are
($a=|a|\exp(i\pi/4)$)
\begin{equation}
{\rm Im}~\Pi(a) \simeq \frac{3\pi}{8\sqrt{2}|a|} +
\frac{\pi}{8}\left(1-4\ln2 \right) - \frac{\pi^3|a|}{48\sqrt{2}}
\label{A.12}
\end{equation}

\setcounter{equation}{0}

\section{Appendix}

Here we consider the asymptotic behavior of the radiation
integral characteristics.
The integral intensity (the radiation length) can be presented as
(see (\ref{3.1}) and (\ref{3.2})):
\begin{eqnarray}
&&P(\beta)=\int_{0}^{1}\frac{xdx}{1-x} \int_{0}^{\infty}dz
\exp(-\beta \eta z)
\Bigg[x^2 \left(\frac{1}{\sinh z}-\frac{1}{z} \right)
\nonumber \\
&&+\frac{1}{\beta \eta}
\left(1+(1-x)^2 \right)
\left( \frac{1}{\sinh^2z}- \frac{1}{z^2}\right) \Bigg],\quad
\eta=\sqrt{\frac{x}{1-x}}
\label{B.1}
\end{eqnarray}
Integrating by parts (over $z$) the second term of the integrand
($\propto 1/(\beta \eta)$) in (\ref{B.1}) we have
\begin{eqnarray}
&& P(\beta)=-\frac{1}{\beta}\int_{0}^{1} \sqrt{\frac{x}{1-x}}
\left(1+(1-x)^2 \right)dx
+\int_{0}^{1}\frac{xdx}{1-x} \int_{0}^{\infty}dz \exp(-\beta \eta z)
\nonumber \\
&&\times \Bigg[x^2\left(\frac{1}{\sinh z}-\frac{1}{z}\right)
+\left(1+(1-x)^2 \right)
\left(1-\coth z+ \frac{1}{z} \right) \Bigg]
\nonumber \\
&&= -\frac{9\pi}{16 \beta}+P_1(\beta)+P_2(\beta),
\label{B.2}
\end{eqnarray}
where
\begin{eqnarray}
&& P_1(\beta)=\int_{0}^{1}\frac{x^3dx}{1-x} \int_{0}^{\infty}dz
\exp(-\beta \eta z)\left(1-\coth z+ \frac{1}{\sinh z} \right),
\nonumber \\
&& P_2(\beta)=2\int_{0}^{1}xdx \int_{0}^{\infty}dz
\exp(-\beta \eta z)\left(1-\coth z+ \frac{1}{z} \right),
\label{B.3}
\end{eqnarray}
We split the function $P(\beta)$ into two functions:
\begin{eqnarray}
\hspace{-10mm}&& P_1(\beta)=P_{11}(\beta)+P_{12}(\beta),
\nonumber \\
\hspace{-10mm}&&P_{11}(\beta)
=\int_{0}^{1}\frac{dx}{1-x} \int_{0}^{\infty}dz
\exp(-\beta \eta z)\left(1-\coth z+ \frac{1}{\sinh z} \right),
\nonumber \\
\hspace{-10mm}&& \displaystyle{P_{12}(\beta)=-
\int_{0}^{1}(1+x+x^2)dx \int_{0}^{\infty}dz
\exp(-\beta \eta z)\left(1-\coth z+ \frac{1}{\sinh z} \right)}.
\label{B.4}
\end{eqnarray}
Transposing the integration order in $P{11}(\beta)$ we get the integral
over $x$
\begin{eqnarray}
\hspace{-10mm}&&\int_{0}^{1}\frac{dx}{1-x}
\exp(-\beta \eta z)=2\int_{0}^{\infty}\frac{ydy}{\beta^2z^2+y^2}\exp(-y)
\nonumber \\
\hspace{-10mm}&& = -2\ln(\beta z) + \int_{0}^{\infty}\ln(\beta^2z^2+y^2)
\exp(-y)dy
\label{B.5}
\end{eqnarray}
In the limit $|\beta| \ll 1$ we have discarding terms $\sim \beta^2$
\begin{equation}
P_{11}(\beta)=-2\int_{0}^{\infty}dz (\ln (\beta z) + C)
\left(1-\coth z+ \frac{1}{\sinh z} \right)
\label{B.6}
\end{equation}
We use Eq.(\ref{A.3}) in the calculation of the integral over $z$ in
$P_{12}(\beta)$
\begin{eqnarray}
&&\int_{0}^{\infty}dz \exp(-\beta \eta z)\left(1-\coth z+
\frac{1}{\sinh z} \right)=2\sum_{k=1}^{\infty}
\left(\frac{1}{2k-1+\beta \eta}-\frac{1}{2k+\beta \eta} \right)
\nonumber \\
&& \simeq 2\ln 2 - \beta \eta \zeta(2).
\label{B.7}
\end{eqnarray}
Taking into account that $\beta=|\beta|\exp (i\pi/4)$ we have
for ${\rm Im}P(\beta)$ at $|\beta| \ll 1$
\begin{eqnarray}
&&{\rm Im}P_{11}(\beta) = -\frac{\pi}{2}\int_{0}^{\infty}dz \left(1-\coth z+
\frac{1}{\sinh z} \right)= -\pi \ln 2,
\nonumber \\
&& {\rm Im}P_{12}(\beta)=\frac{|\beta| \zeta(2)}{\sqrt{2}}
\int_{0}^{1} \sqrt{\frac{x}{1-x}} (1+x+x^2)dx= \frac{19 \pi}{16\sqrt{2}}
|\beta| \zeta(2).
\label{B.8}
\end{eqnarray}
We will use Eqs.(\ref{A.4})-(\ref{A.6}) in the calculation of the integral
over $z$ in $P_{2}(\beta)$ Eq.(\ref{B.3})
\begin{eqnarray}
&& \int_{0}^{\infty}dz\exp(-\beta \eta z)\left(1-\coth z+ \frac{1}{z} \right)
=\ln \frac{2}{\beta \eta}-C + \sum_{k=1}^{\infty}
\frac{\beta \eta}{k(2k+\beta \eta)}
\nonumber \\
&&\simeq \ln \frac{2}{\beta \eta}-C+\frac{1}{2}\beta \eta\zeta(2)
\label{B.9}
\end{eqnarray}
As a result we get
\begin{equation}
{\rm Im}P_{2}(\beta)=2\int_{0}^{1}x dx \left(-\frac{\pi}{4}+
\frac{|\beta|}{2\sqrt{2}}\sqrt{\frac{x}{1-x}}\zeta(2) \right)
=-\frac{\pi}{4} +\frac{3\pi |\beta|}{8\sqrt{2}} \zeta(2)
\label{B.10}
\end{equation}
Substituting (\ref{B.8}) and (\ref{B.10}) into (\ref{B.2}) we obtain
\begin{equation}
{\rm Im}P(\beta)= \frac{9\pi}{16 \sqrt{2}|\beta|} - \pi \ln 2
-\frac{\pi}{4} +\frac{25\pi |\beta|}{16\sqrt{2}} \zeta(2) +O(\beta^2)
\label{B.11}
\end{equation}

In the calculation of the total probability of radiation one have
to make the substitution in (\ref{B.1})
\[
\int_{0}^{1}\frac{xdx}{1-x}\ldots~\rightarrow~
\int_{0}^{1}\frac{dx}{1-x}\dots.
\]
Then
\begin{equation}
T(\beta)=-\frac{1}{\beta}\int_{0}^{1}\frac{dx}{\sqrt{x(1-x)}}
(1+(1-x)^2)+t_1(\beta)+t_2(\beta),
\label{B.12}
\end{equation}
where
\begin{eqnarray}
\hspace{-6mm}&&{\rm Im}~t_{1}(\beta)
= -\pi \ln 2 +\frac{1}{\sqrt{2}}|\beta| \zeta(2)
\int_{0}^{1}\sqrt{\frac{x}{1-x}}(1+x)dx= -\pi \ln 2
+\frac{7\pi}{8\sqrt{2}}|\beta| \zeta(2),
\nonumber \\
\hspace{-6mm}&&{\rm Im}~t_{2}(\beta)
= -\frac{\pi}{2} +\frac{1}{2\sqrt{2}}|\beta| \zeta(2)
\int_{0}^{1}\sqrt{\frac{x}{1-x}}dx= -\frac{\pi}{2}
+\frac{\pi}{2\sqrt{2}}|\beta| \zeta(2),
\nonumber \\
\hspace{-6mm}&&{\rm Im}~T(\beta)
= \frac{11\pi}{8\sqrt{2}|\beta|} -\frac{\pi}{2} -\pi \ln 2
+\frac{11\pi}{8\sqrt{2}}|\beta| \zeta(2) +O(\beta^2)
\label{B.13}
\end{eqnarray}

\setcounter{equation}{0}

\section{Appendix}

Here we consider the asymptotic behavior in the
region where the LPM effect is
weak. In this region in Eq.(\ref{A.1}) the parameter $|a| \gg 1$
and in the integral over $z$ the interval $z \ll 1$ contributes,
and we can use expansions
\begin{eqnarray}
&&\frac{1}{\sinh z}-\frac{1}{z}=-\frac{z}{6}+\frac{7z^3}{360}
-\frac{31z^5}{15120} + \ldots,
\nonumber \\
&&\frac{1}{\sinh^2 z}-\frac{1}{z^2}= -\frac{1}{3}+ \frac{z^2}{15}-
\frac{2z^4}{189} + \ldots
\label{C.1}
\end{eqnarray}
Substituting these expansions into (\ref{A.1}) and taking into
account that $a=|a|\exp(i\pi/4)$ we get
\begin{eqnarray}
&& {\rm Im}~\Pi(a) = \int_{0}^{\infty} d\xi \Bigg[\frac{1}{6|a|^2 \cosh^2 \xi}
- \frac{31}{126|a|^6 \cosh^6 \xi} +
\left(1-\frac{1}{2 \cosh^2 \xi} \right)
\nonumber \\
&&\times \Bigg(\frac{1}{3|a|^2 \cosh^2 \xi}
 -\frac{16}{63|a|^6 \cosh^6 \xi}\Bigg)\Bigg]=
\frac{7}{18|a|^2}\left(1-\frac{184}{343|a|^4}\right).
\label{C.2}
\end{eqnarray}

We turn to ${\rm Im}~T(\beta)$ Eq.(\ref{B.1}) at
$|\beta| \gg 1 (\beta = |\beta|\exp(i\pi/4))$. The integral over $z$
coincides with this integral in (\ref{C.1}). The integral over $x$
gives for $x \sim 1$ the same structure as in (\ref{C.2}):
the main term $\sim 1/|\beta|^2$ and the correction $\sim 1/|\beta|^6$.
In the region $x \sim 1/|\beta|^2$ where the influence of the LPM effect
is significant, the correction is proportional to the phase space:
$\int_{}^{}xdx  \sim 1/|\beta|^2$. This is the main correction.
Because of this we split the integration interval over $x$ into two
intervals: 1) $0 \leq x \leq x_0$ and 2) $x_0 \leq x \leq 1$,
where $1/|\beta|^2 \ll x_0 \ll 1$.
In the first interval
\begin{equation}
P_1(\beta, x_0) \simeq \frac{2}{\beta}\int_{0}^{x_0}\frac{xdx}{\eta (1-x)}
\int_{0}^{\infty}
\exp(-\beta \eta z)
\left( \frac{1}{\sinh^2z}- \frac{1}{z^2}\right)dz.
\label{C.3}
\end{equation}
Integrating over $z$ by part and then over $x$ we have
\begin{equation}
P_1(\beta, x_0) \simeq \frac{2x_0}{3\beta^2} + \frac{8}{\beta^4}
\int_{0}^{\infty} \frac{dz}{z^2}\left(
1-(\beta z \sqrt{x_0}+1) \exp(-\beta z \sqrt{x_0}) \right)
\left( \frac{\cosh z}{\sinh^3z}- \frac{1}{z^3}\right).
\label{C.4}
\end{equation}
In the integral which contains the term $\beta z \sqrt{x_0}$ the interval
$z \ll 1$ contributes so that
\begin{equation}
\int_{0}^{\infty} \frac{dz}{z^2}
\beta z \sqrt{x_0} \exp(-\beta z \sqrt{x_0})
\left(-\frac{z}{15}\right) = \frac{1}{15}
\label{C.5}
\end{equation}
In the remaining integral we split the interval of integration into two:
1) $0 \leq z \leq z_0$, 2)$z_0 \leq z < \infty
~(1/(\beta \sqrt{x_0}) \ll z_0 \ll 1)$, then
\begin{eqnarray}
&&\int_{0}^{\infty} \frac{dz}{z^2}\left(
1- \exp(-\beta z \sqrt{x_0}) \right)
\left( \frac{\cosh z}{\sinh^3z}- \frac{1}{z^3}\right)
\nonumber \\
&&\simeq \int_{0}^{z_0} \frac{dz}{z^2}\left(
1- \exp(-\beta z \sqrt{x_0}) \right)
\left(-\frac{z}{15}\right)
+\int_{z_0}^{\infty} \frac{dz}{z^2}
\left(\frac{\cosh z}{\sinh^3z}- \frac{1}{z^3}\right)
\nonumber \\
&&\simeq -\frac{1}{15} \left(\ln (-\beta z_0 \sqrt{x_0})+C \right)
+\int_{z_0}^{\infty} \frac{dz}{z^2}
\left(\frac{\cosh z}{\sinh^3z}- \frac{1}{z^3}\right)
\label{C.6}
\end{eqnarray}
So we have
\begin{equation}
{\rm Im}~P_1(\beta, x_0) \simeq \frac{2x_0}{3|\beta|^2} -
\frac{2\pi}{15|\beta|^4}
\label{C.7}
\end{equation}
In the second interval over $x~(1 \geq x \geq x_0)$ the interval
$z \ll 1$ contributes as well as in the first term of $P(\beta)$
Eq.(\ref{B.1}) which we include here
\begin{equation}
{\rm Im}~P_2(\beta, x_0) \simeq \frac{1}{2|\beta|^2} -
\frac{2x_0}{3|\beta|^2} + O(\frac{1}{\beta^6})
\label{C.8}
\end{equation}
Adding (\ref{C.7}) and (\ref{C.8}) we get for $|\beta| \gg 1$
\begin{equation}
{\rm Im}~P(\beta) \simeq \frac{1}{2|\beta|^2} - \frac{2\pi}{15|\beta|^4}
\label{C.9}
\end{equation}

\newpage

{\bf Figure captions}

\vspace{15mm}
\begin{itemize}

\item {\bf Fig.1} The pair creation spectral probability
$\displaystyle{\frac{dW_p}{dx}, x=\frac{\varepsilon}{\omega}}$
in gold in terms of the exact total Bethe-Heitler probability taken
with the Coulomb corrections (see Eq.(\ref{2.23a})).
\begin{itemize}
\item Curve BH is the Bethe-Heitler spectral probability
(see Eq.{2.23});
\item curve T1 is the total contribution (the sum of the
logarithmic approximation $dW_p^c/d\varepsilon$ (\ref{2.10}) and
the first correction to the spectral probability $dW_p^1/d\varepsilon$
(\ref{2.15})) for the photon energy $\omega=2.5$~TeV;
\item curve 2 is the is the logarithmic approximation
$dW_p^c/d\varepsilon$ (\ref{2.10}), curve c2 is the first correction
to the spectral probability $dW_p^1/d\varepsilon$
(\ref{2.15})) and curve T2 is the
sum of the previous contributions for the photon energy $\omega=25$~TeV;
\item curves 3, c3, T3 are the same for the photon energy $\omega=250$~TeV;
\item curves 4, c4, T4 are the same for the photon energy $\omega=2500$~TeV;
\end{itemize}

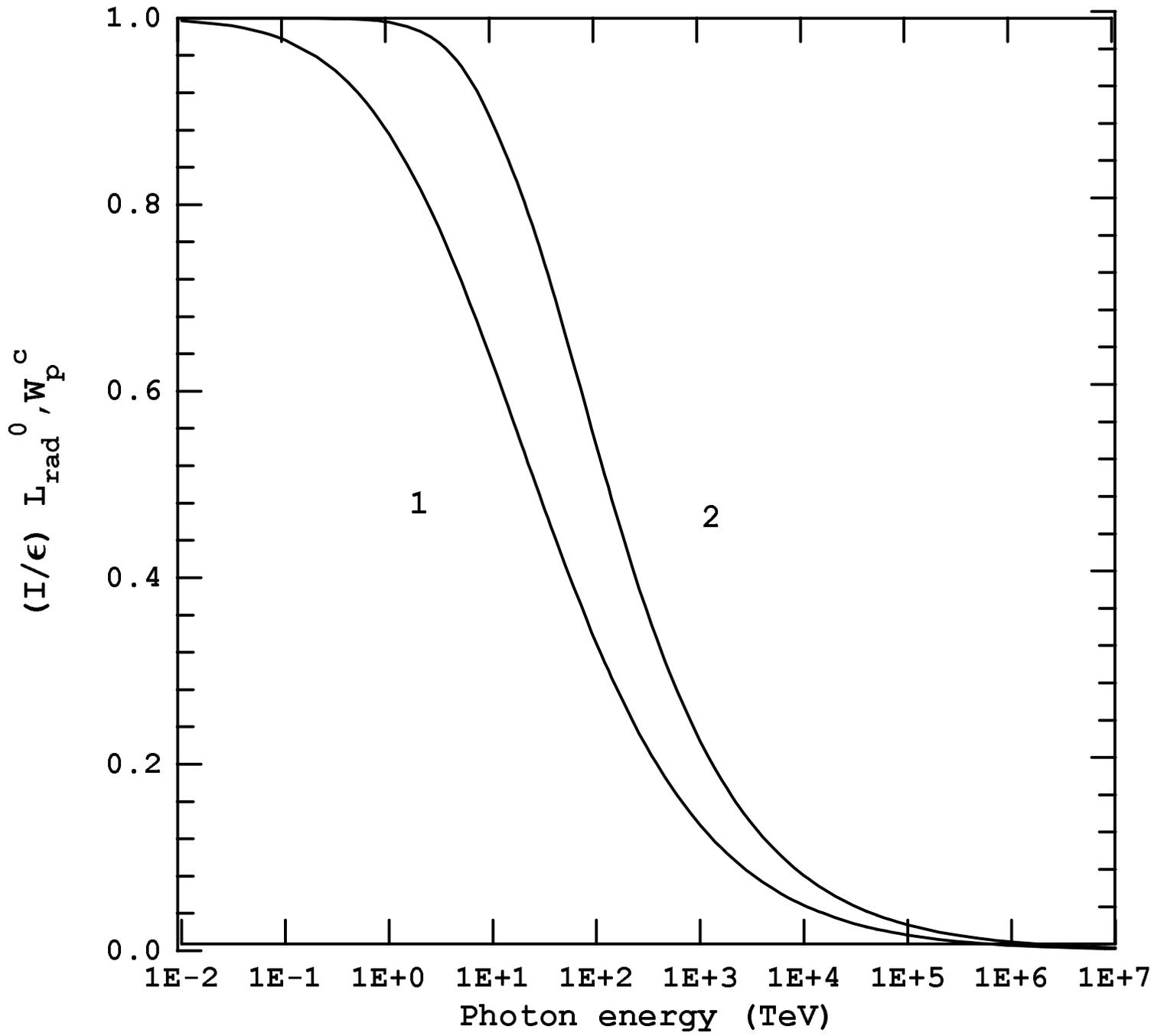
\item {\bf Fig.2} The relative energy losses of electron
per unit time in terms
of the Bethe-Heitler radiation length $L_{rad}^0$:
$\displaystyle{\frac{I}{\varepsilon}L_{rad}^0}$ in gold vs the initial energy
of electron
(curve 1) and the total pair creation probability
per unit time $W_p^c$ (see Eq.(\ref{2.23b}))in terms
of the Bethe-Heitler total probability of pair creation $W_{p0}^{BH}$
(see Eq.(\ref{2.23a}))
in gold vs the initial energy of photon (curve 2).

\item {\bf Fig.3} The spectral intensity of radiation
$\displaystyle{\omega \frac{dW}{d\omega}= x\frac{dW}{dx},~x=
\frac{\omega}{\varepsilon}}$
in gold in terms of $3 L_{rad}$ taken
with the Coulomb corrections (see Eq.(\ref{3.8})).
\begin{itemize}
\item Curve BH is the Bethe-Heitler spectral intensity
(see Eq.{3.6});
\item curve 1 is the is the logarithmic approximation
$\omega dW_c/d\omega$ Eq.(2.28) of \cite{L1}, curve c1
is the first correction
to the spectral intensity $\omega dW_1/d\omega$ Eq.(2.33) of \cite{L1}
and curve T1 is the
sum of the previous contributions for the electron energy
$\varepsilon=25$~GeV;
\item curve 2 is the is the logarithmic approximation
$\omega dW_c/d\omega$ Eq.(2.28) of \cite{L1}, curve c2
is the first correction
to the spectral intensity $\omega dW_1/d\omega$ Eq.(2.33) of \cite{L1}
and curve T2 is the
sum of the previous contributions for the electron energy
$\varepsilon=250$~GeV;
\item curves 3, c3, T3 are the same for the electron energy
$\varepsilon=2.5$~TeV;
\item curves 4, c4, T4 are the same for the electron energy
$\varepsilon=25$~TeV;
\end{itemize}

\end{itemize}

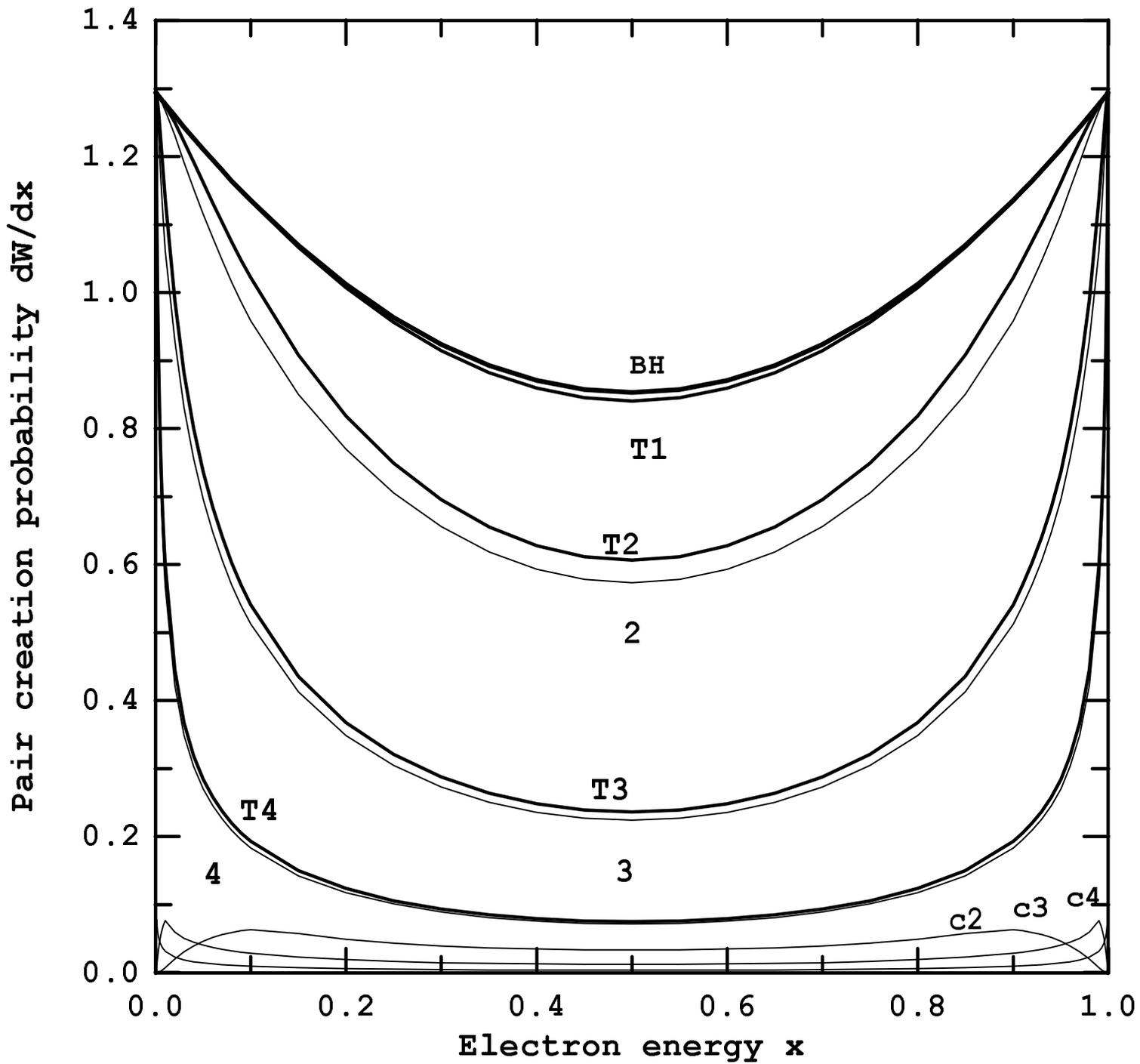
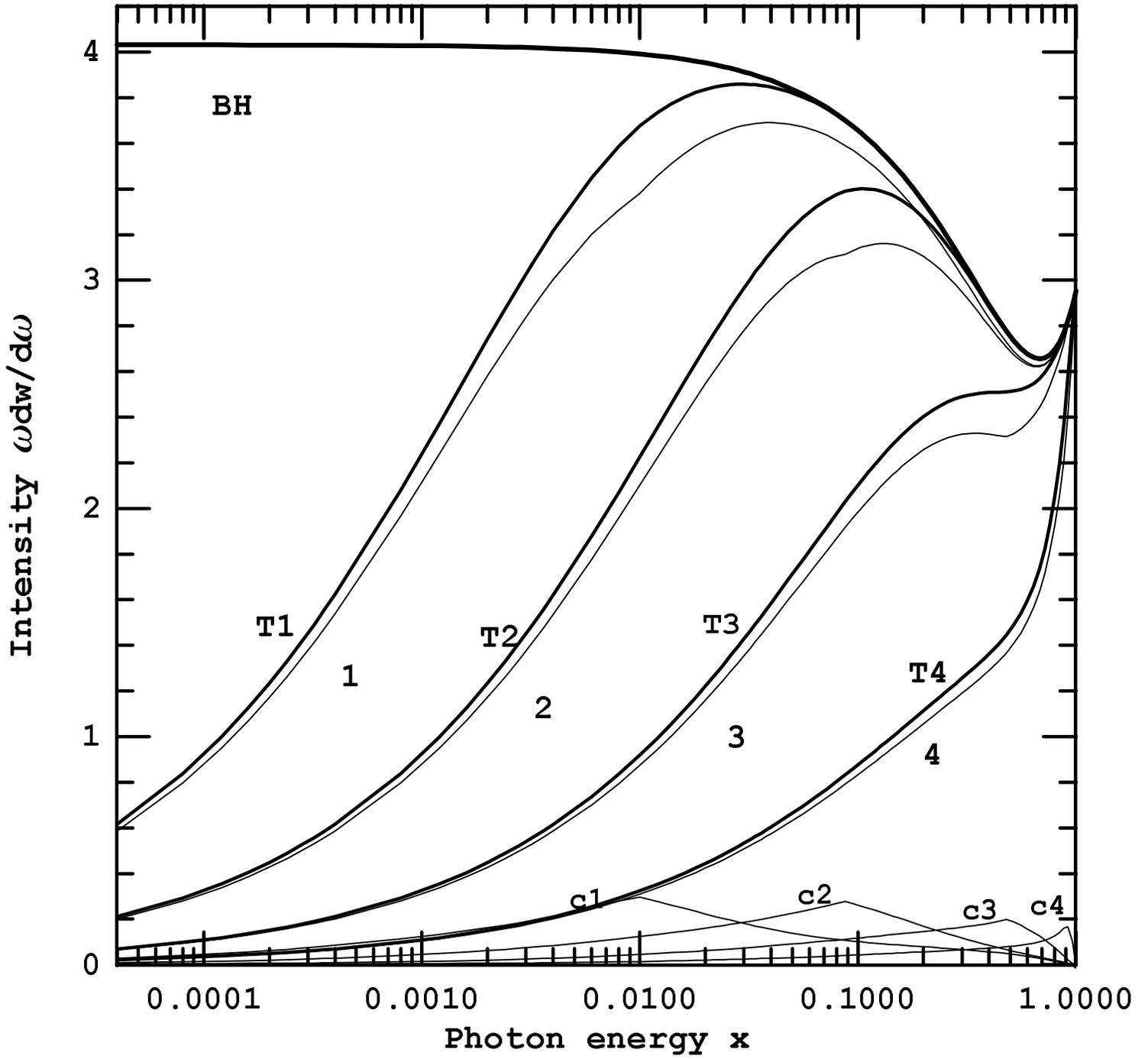
\end{document}